\documentclass[a4paper, 11pt]{article}
\pdfoutput=1 
\usepackage{jcappub}

\title{A realistic assessment of the CTA sensitivity to dark matter annihilation}

\author[a]{Hamish Silverwood,}
\author[a]{Christoph Weniger,}
\author[b]{Pat Scott}
\author[a]{and Gianfranco Bertone}

\affiliation[a]{GRAPPA, University of Amsterdam\\Science Park 904, 1098 XH
Amsterdam, The Netherlands}
\affiliation[b]{Department of Physics, Imperial College London,\\Blackett
Laboratory, Prince Consort Road, London SW7 2AZ, United Kingdom}

\emailAdd{h.g.m.silverwood@uva.nl}
\emailAdd{c.weniger@uva.nl}
\emailAdd{p.scott@imperial.ac.uk}
\emailAdd{gf.bertone@gmail.com}

\newcommand{\Fermi}{\textit{Fermi}}
\newcommand{\ie}{i.e.~}
\newcommand{\diff}{\mathrm{d}}
\newcommand{\modelsym}{\mu}
\newcommand{\datasym}{n}
\newcommand{\model}{{\boldsymbol\modelsym}}
\newcommand{\modelij}{{\modelsym_{ij}}}
\newcommand{\data}{{\boldsymbol{\datasym}}}
\newcommand{\dataij}{{\datasym_{ij}}}
\newcommand{\alphaij}{{\alpha_{ij}}}
\newcommand{\betai}{{\beta_{i}}}

\abstract{We estimate the sensitivity of the upcoming CTA gamma-ray observatory
to DM annihilation at the Galactic centre, improving on previous analyses in a
number of significant ways.  First, we perform a detailed analyses of all
backgrounds, including diffuse astrophysical emission for the first time in a
study of this type. Second, we present a statistical framework for including
systematic errors and estimate the consequent degradation in sensitivity.
These errors may come from e.g.~event reconstruction, Monte Carlo determination
of the effective area or uncertainty in atmospheric conditions. Third, we show
that performing the analysis on a set of suitably optimised regions of interest
makes it possible to partially compensate for the degradation in sensitivity
caused by systematics and diffuse emission. To probe dark matter with the
canonical thermal annihilation cross-section, CTA systematics like non-uniform
variations in acceptance over a single field of view must be kept below the
0.3\% level, unless the dark matter density rises more steeply in the centre of
the Galaxy than predicted by a typical Navarro-Frenk-White or Einasto profile.
For a contracted $r^{-1.3}$ profile, and systematics at the 1\% level, CTA can
probe annihilation to $b\bar{b}$ at the canonical thermal level for dark matter
masses between 100 GeV and 10 TeV.}

\begin{document}
\maketitle

\section{Introduction}
\label{sec:Intro}

Current and upcoming gamma-ray experiments are ideally suited to probing the
nature of dark matter (DM), by searching for high-energy photons from
annihilation of Weakly Interacting Massive Particles (WIMPs;
\cite{Jungman96,Bergstrom00,Bertone05,BringmannWeniger,BertoneBook}).  Other
so-called {\it indirect} DM detection strategies include searches for neutrinos
\cite{SuperK11,IC40DM,IC79,IC22Methods,Silverwood12,Rott13} or anti-matter
\cite{Salati10,PAMELAantiproton,Dal12,Bergstrom13,Fornengo13,Carlson14,Cirelli:2014qia}
produced by DM annihilation.  Indirect detection is particularly appealing at
the present time, owing to a lack of convincing evidence from DM {\it direct}
detection \cite{CerdenoGreen10,Pato11,Strege12,SavageRev,XENON2013,LUX13},
which seeks to measure energy exchanged in collisions between nuclei and DM in
underground experiments, and {\it collider} searches looking for new particles
at the Large Hadron Collider
\cite{Goodman10,Fox12b,CMS12a,ATLASDM14,BertoneBook}.

Recent data from the Large Area Telescope (LAT) aboard the \textit{Fermi}
satellite set stringent and robust limits on the annihilation cross section of
WIMPs as a function of the DM particle mass, based on a lack of excess
gamma-ray emission from dwarf spheroidal galaxies \cite{Ackermann:2013yva}.
\textit{Fermi} data have also led to the discovery of excess emission from the
Galactic centre (GC), which has been interpreted in terms of DM annihilation
(see e.g.~Ref.~\cite{Daylan:2014rsa} and references therein). Furthermore, the
Imaging Air Cherenkov Telescopes (IACTs) HESS, VERITAS and MAGIC search for DM
signals in dwarf spheroidal galaxies and the GC~\cite{Abramowski:2010aa,
Aleksic:2013xea, AlexGeringer-SamethfortheVERITAS:2013fra}, with the current
strongest limits on TeV-scale DM coming from HESS observations of the
GC~\cite{Abramowski:2011hc}.

One of the next major steps in high-energy gamma-ray astrophysics will be the
construction of the Cherenkov Telescope Array (CTA), which is currently in the
design phase~\cite{Consortium:2010bc} and expected to start operations in 2019.
Several estimates of the sensitivity of CTA to gamma-rays from DM annihilation
exist in the literature. All of these agree on the fact that CTA will improve
existing constraints for values of the DM particle mass above
$\cal{O}$(100)\,GeV, but substantial differences exist, up to one order of
magnitude or more in annihilation cross section for a given mass, depending on
the assumptions made about the telescope array configuration, analysis setup
and observation time  \cite{Doro:2012xx,Wood:2013taa, Pierre14}. 

In this paper we carry out a new estimation of the sensitivity of CTA to DM
annihilation, and compare this to the sensitivity of \textit{Fermi}.  We
improve on previous analyses in a number of ways. First, we present a detailed
discussion of all backgrounds, including cosmic-ray protons and electrons, and
for the first time in the context of CTA and DM, the effects of the diffuse
astrophysical emission.  We estimate this from \textit{Fermi}-LAT data,
suitably extrapolated above 500\,GeV in order to cover the range of energies
relevant for CTA.  Second, we study the impact of systematic errors.  There are
many sources of systematic uncertainty inherent in CTA measurements: event
reconstruction, Monte Carlo determination of the effective area, and
uncertainty in atmospheric conditions \cite{Consortium:2010bc}. A detailed
assessment can only realistically be performed by the CTA Collaboration itself
after the instrument is built; here we instead present a simple but
comprehensive statistical framework with which the impacts of these systematics
on the sensitivity of CTA to DM annihilation can be illustrated and evaluated.
Third, we show that performing the analysis over a series of suitably optimised
regions of interest (RoIs) partially compensates for the degradation in
sensitivity due to systematics and backgrounds. We carry out a multi-RoI
`morphological' analysis of the gamma-ray emission, and demonstrate how it
improves the CTA sensitivity to DM compared to the so-called `Ring' method
previously discussed in the literature~\cite{Doro:2012xx}.

The paper is organised as follows:  in Sec.\ \ref{sec:Experiments} we describe
the CTA and \textit{Fermi}-LAT experiments; in Sec.\
\ref{sec:DarkMatterSignals} we review the basics of indirect detection with
gamma-rays; in Sec.\  \ref{sec:Backgrounds} we discuss and quantify the
cosmic-ray and diffuse gamma-ray background for CTA; in Sec.\
\ref{sec:Analysis} we present our analysis strategy, in particular the
implementation of systematic errors and the RoIs relevant for our analysis; we
present our results in Sec.\ \ref{sec:Results} and conclusions in Sec.\
\ref{sec:conclusion}.

\section{The CTA and \textit{Fermi}-LAT experiments}
\label{sec:Experiments}

\subsection{The CTA and other imaging air Cherenkov telescopes}
\label{iacts}

Gamma rays in the GeV to TeV regime initiate electromagnetic cascades in the
atmosphere, which start at an altitude of 10--20\,km and generate a focused
cone of Cherenkov light that typically covers several hundred meters on the
ground.  Air Cherenkov telescopes detect gamma rays by observing this dim
Cherenkov light with optical telescopes.  The overall light yield, the shape
and the orientation of the air shower gives information about the energy and
arrival direction of the gamma ray.  Because observations can only be performed
during (nearly) moonless nights, the observation time per year is limited to
approximately 1000 hours.

Current active Imaging Air Cherenkov Telescopes (IACTs) are HESS II (Namibia;
\cite{Aharonian:2006pe}), VERITAS (Arizona; \cite{Holder:2006gi}), and MAGIC
(La Palma\footnote{\url{http://magic.mpp.mpg.de/}}). Although all three
instruments have an active program to search for DM signals in various regions
of the sky (see e.g.~Refs.~\cite{Abramowski:2010aa, Aleksic:2013xea,
AlexGeringer-SamethfortheVERITAS:2013fra}), only HESS can observe the GC high
above the horizon ($\theta_z\sim 6^\circ$, compared to
$\theta_z\gtrsim57^\circ$ for VERITAS and MAGIC).  As a shorter transmission
length through the atmosphere allows for a lower threshold energy, HESS is
ideally situated to search for DM signals from the GC.  Non-detection of a DM
signal by HESS provides the strongest current limits on the DM
self-annihilation cross section, for DM masses around the TeV
scale~\cite{Abramowski:2011hc}.

CTA will consist of several tens of telescopes of at least 3 different types,
with sizes between about 5 and 24 meters, covering an area of several square
kilometres.  The sensitivity will be a factor ten better than current
instruments, the field of view (FoV) will be up to 10$^\circ$ in diameter, and
the energy threshold $\mathcal{O}$(10\,GeV).

CTA is envisaged as a two-part observatory, with southern and northern sites.
CTA South is the most relevant for DM searches towards the Galactic centre.
The location of the southern array has not yet been settled, but the main
candidates are now the Khomas Highlands in Namibia and Cerro Paranal in Chile.
The final design is also not yet fixed.  Apart from construction and
maintenance questions, relevant remaining design choices are the relative
emphasis on higher or lower energies, the angular and energy resolution, and
the FoV.  A first detailed Monte Carlo (MC) analysis was presented in
Ref.~\cite{Bernlohr:2012we}, where 11 different configurations for the southern
array were discussed.  Depending on the array configuration and gamma-ray
energy, the point source sensitivity varies within a factor of
five,\footnote{This is based on the adoption of a standard Hillas-based
analysis.  This is a classical analysis method, based on zeroth (amplitude),
first (position) and second (width and orientation) momenta of the
images~\cite{Hillas85}.} and can be further improved by about a factor of two
with alternative analysis methods \cite{Bernlohr:2012we}.

In this paper we will concentrate on the proposed configuration known as `Array
I'~\cite{Bernlohr:2012we},\footnote{In particular, we adopt the version based
on the Hillas-parameter analysis of the MPIK group. The choice of analysis
method can impact the projected sensitivity; e.g.~the Paris-MVA method produces
an effective area that is at energies around 1\,TeV about $\sim7$ times that of
the MPIK method \cite{Bernlohr:2012we}, making Paris-MVA derived limits up to
$\sim 2.6$ stronger than those using MPIK. However, this only holds for
background limited observations.  Observations of the diffuse emission are
actually systematics limited in most of the parameter range of interest,
which makes differences in the effective area much less relevant.  For this
reason, we concentrate here on the better documented MPIK method.} 
which is a balanced configuration with three
large ($\sim$24\,m aperture), 18 medium ($\sim$12\,m) and 56 small telescopes
($\sim$4--7\,m).   This configuration provides a good compromise in sensitivity
between low and high energies.  One advantage of using Array I is that
extensive information on the effective area, angular resolution, energy
resolution and background rates is available.  Furthermore, a very similar
array has been used in previous DM sensitivity studies: Array E in Ref.\
\cite{Doro:2012xx}, and the Paris-MVA version of Array I in Ref.\
\cite{Pierre14}. The point-source sensitivities of Arrays E and I agree very
well at energies $\lesssim1$\,TeV, whereas at higher energies Array I is more
sensitive (but only by a factor of less than two).

For convenience, here we summarise some of the main performance aspects of
Array I:  it features an effective area of about 100\,m$^2$ at its threshold
energy of 20\,GeV, which then increases quickly with energy to about
$4\times10^5\,\rm m^2$ at 1\,TeV and $3\times10^6\,\rm m^2$ at 10\,TeV.  The
angular resolution in terms of the $68\%$ containment radius is about
$r_{68}\simeq0.3^\circ$ at threshold, and drops to below $0.06^\circ$ at
energies above 1\,TeV.  The energy resolution is relatively large at threshold,
with $\sigma(E)/E\simeq 50\%$, but drops to below $10\%$ at energies above
1\,TeV.

As we will discuss below in more detail, the large effective area, irreducible
background of cosmic-ray (CR) electrons, and the fact that electrons look
identical to photons for an IACT, mean that the number of events passing
analysis cuts will be extremely high, especially at low energies and for
analyses of diffuse and extended sources.  Statistical errors will therefore be
small, making systematic uncertainties in the event identification particularly
important.  In fact, in the case of searches for DM particles with GeV to TeV
masses, these systematics will turn out to be one of the limiting factors.
Detailed instrumental uncertainties (e.g.~the covariance matrix of
reconstruction efficiencies in different regions of the FoV) will probably only
become available once the instrument starts to operate, so in this paper we
resort to a few well-motivated benchmark scenarios.

The traditional observing strategy employed by IACTs in searching for DM
annihilation (e.g.~Ref.~\cite{Abramowski:2011hc}) involves defining two regions
on the sky expected to have approximately the same regular astrophysical
emission, but different amounts of DM annihilation.  The region with the larger
expected annihilation is dubbed the `ON' region, the other is called the `OFF'
region, and the analysis is performed using a test statistic defined as the
difference in photon counts from the two regions.  This is referred to as an
`ON-OFF' analysis, and obviously obtains the most power when the ON and OFF
RoIs are chosen to differ as much as possible in their predicted annihilation
rates.

The RoIs chosen for ON-OFF analyses may lie in the same or very different FoVs.
Different FoVs allow a greater contrast in DM signal between ON and OFF
regions, but have the potential to introduce differential systematics across
the two FoVs.  The `Ring method' \cite{Doro:2012xx} is an ON-OFF analysis
technique optimised for DM searches towards the GC with IACTs, which fits the
ON and OFF regions into a single FoV, producing an approximately constant
acceptance across the entire analysis region.  Although both regions are
expected to contain DM and background contributions, in the Ring method the ON
and OFF regions are typically referred to as the `signal' and `background'
regions.  For simplicity, here we just call them ON and OFF.

\begin{figure}
    \begin{center}
        \includegraphics[width=0.49\linewidth]{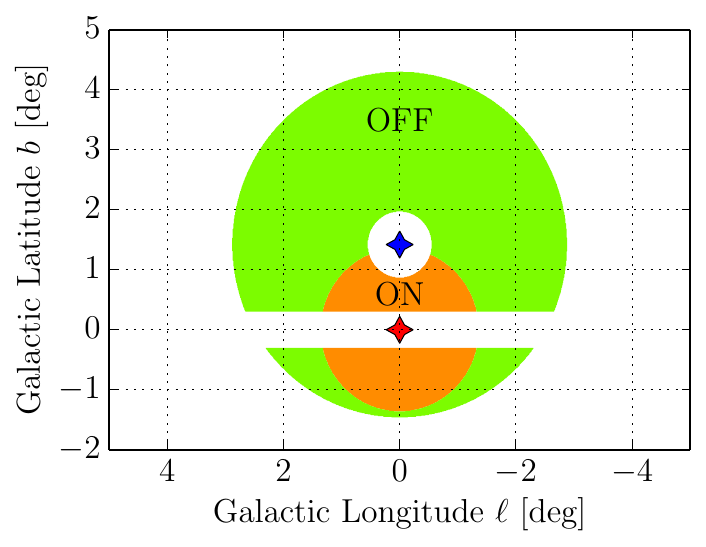}
        \includegraphics[width=0.49\linewidth]{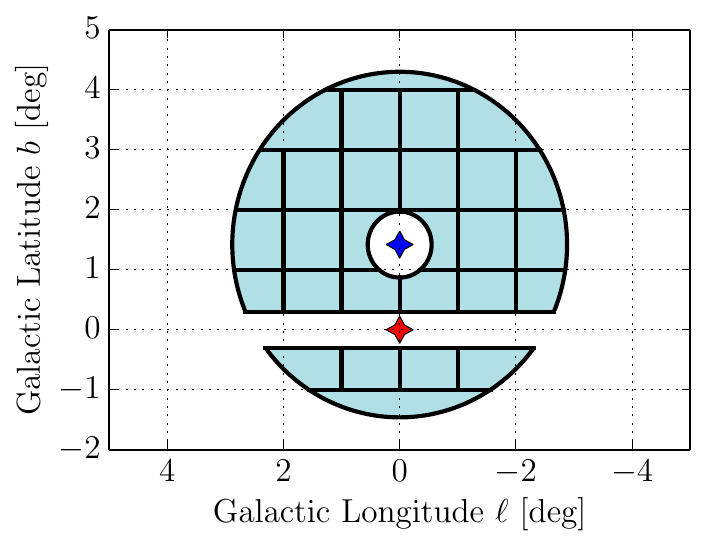}
    \end{center}
    \caption{The different RoIs that we consider in this paper. The red star indicates the GC, while the blue star indicates the centre of the FoV. \emph{Left:}
    RoIs used in the Ring method of Ref.~\cite{Doro:2012xx} as `signal' and
    'background' regions; we refer to these as simply `ON' and `OFF' regions,
    respectively. A diagram detailing the construction of the annulus and the ON-OFF RoIs is given in Ref.~\cite{Doro:2012xx}, and the exact dimensions we use are given in Subsection \ref{subsec:RegionsAndJfactors}. \emph{Right:} Separation of the ON and OFF RoIs into 28
    sub-RoIs, which we use in our morphological analysis.}
    \label{fig:rois}
\end{figure}

A simple way to model the results of an ON-OFF analysis is to construct a
Skellam likelihood \cite{Skellam,Cline13b,Pierre14}, which is based on the
expected difference between two Poisson counts (\ie in the ON and OFF regions).
However, once the assumption that astrophysical backgrounds are identical in
the ON and OFF regions becomes questionable, a more straightforward method is
simply to carry out a regular binned likelihood analysis.  In this case, one
predicts the photon counts in each RoI using detailed background and signal
models, and compares them directly to the absolute number of photons observed
in each RoI.  This is the strategy that we investigate here for CTA, using both
the original Ring method RoIs and a finer spatial binning.  We show these two
sets of RoIs in Fig.~\ref{fig:rois}, and discuss their optimisation in Sec.\
\ref{sec:Analysis}. We still refer to the two-RoI analysis as the `Ring method'
even though we carry out a full likelihood analysis rather than an ON-OFF
analysis.  We refer to the multi-RoI analysis as a `morphological analysis', as
it uses the expected spatial distribution of the DM signal to improve limits.

\subsection{The Fermi Large Area Telescope}
\label{fermi}

The LAT aboard the \textit{Fermi} satellite is a pair-conversion detector, with
a large FoV and an energy range from 30\,MeV to above 300\,GeV.  Since its
launch in 2008 the dominant observation strategy has been a full-sky survey
providing roughly equal exposure in all directions of the sky.  The
\textit{Fermi}-LAT is a formidable tool in the search for signals of DM
annihilation.  It currently provides the strongest constraints on the DM
self-annihilation cross section, based on a combined observation of 15 dwarf
spheroidal galaxies~\cite{Ackermann:2013yva}.  

The \textit{Fermi}-LAT provides an accurate measurement of the Galactic diffuse
emission (GDE) up to energies of about 100\,GeV.  Above 100\,GeV the number of
detected photons becomes very small, due to the LAT's comparatively small
effective area ($\sim8000\rm\ cm^2$).  This is about the energy where the
acceptance of IACTs starts to become sizable.  Close to the Galactic disc and
the GC, LAT measurements are actually dominated by GDE. This diffuse emission
will be a critical foreground for any DM search, and---depending on the details
of the search strategy---might mimic a DM signal.

Compared to IACTs, the LAT is an extremely clean gamma-ray telescope.  Its
plastic scintillator anti-coincidence detector, together with cuts on different
event reconstruction quality estimators, allows proton and electron
contamination in the gamma-ray sample to be suppressed to a level well below
the very dim extragalactic gamma-ray background.  The main uncertainties in the
measured fluxes, especially at the high energies we are interested in, are
hence of a statistical nature.  The systematic uncertainty of the effective
area is at the level of $10\%$~\cite{Ackermann:2012kna}, so we will neglect it
throughout.

\section{The dark matter signal}

\label{sec:DarkMatterSignals}

Assuming that the annihilation cross-section of DM does not depend on the
relative velocity between particles, the calculation of the gamma-ray flux from
DM annihilation can be divided into two parts.

\smallskip

The first part accounts for the DM distribution, and is commonly known as the
`$J$-factor'.  This is an integral of the DM density squared along the lines of
sight (l.o.s.) within a cone $\Delta\Omega$ that covers a certain RoI:
\begin{equation}
    J\left(\Delta\Omega\right) = \int_{\Delta\Omega} \text{d}\Omega
    \int_\text{l.o.s} \text{d}l \, \rho_\text{DM} (r)^2 \label{Jfactor}.
\end{equation}
The DM density profiles seen in $N$-body simulations of Milky-Way type galaxies
are best fit by the Einasto profile \cite{nfwsmooth} 
\begin{equation}
    \label{einasto}
    \rho_\text{DM}(r) \propto \exp \left( -\frac{2}{\alpha} \left[
    \left(\frac{r}{r_s}\right)^\alpha -1 \right] \right),
\end{equation}
which we assume for our calculation of the $J$-factors for the GC.  We
normalise to a local DM density of $\rho_\text{DM}(r_\odot) =
0.4\,\text{GeV\,cm}^{-3}$, choosing $\alpha = 0.17$, $r_s = 20$\,kpc, and
$r_\odot = 8.5$\,kpc \cite{Pieri:2009je}.

We will also use a slightly contracted generalised NFW profile, to indicate how
limits would improve with a more optimistic DM distribution.  We parameterise
this profile as
\begin{equation}
    \rho_\text{DM}(r) \propto \frac{1}{r^{\gamma}(r_s + r)^{3-\gamma}}\;,
\end{equation}
where $\gamma=1.3$ is the inner slope of the profile that we will adopt below,
and $r_s=20\,\rm kpc$ the scale radius.  We normalised this profile in the same
way as the Einasto profile (Eq.\ \ref{einasto}).

The second part of the flux calculation covers the actual particle model for
DM.  Together with the $J$-factor, it yields the differential flux of DM signal
photons,
\begin{equation}
    \label{flux_per_E}
    \frac{\text{d} \Phi}{\text{d} E} = \frac{\langle \sigma v \rangle}{8 \pi\,
    m_\text{DM}^2} \frac{\text{d} N_\gamma}{\text{d} E}\,
    J\left(\Delta\Omega\right).
\end{equation}
Here $\langle \sigma v \rangle$ is the velocity-averaged DM self-annihilation
cross section, in the limit of zero relative velocity.  The DM mass is given by
$m_\text{DM}$, and ${\text{d} N_\gamma}/{\text{d} E}$ is the annihilation
spectrum, \ie the average number of photons produced per annihilation per
energy interval, which depends on the particular annihilation channel.  For the
gamma-ray yields, we will take the results from Ref.~\cite{Cirelli:2010xx}.

\smallskip

To calculate the expected number of signal events in a given observing time
$T_\text{obs}$, between energies $E_0$ and $E_1$, we weight
Eq.~\eqref{flux_per_E} by the energy-dependent effective area $A_\text{eff}$
and integrate over the energy range in question,
\begin{equation}
    \mu^\text{DM} = T_\text{obs}\int_{E_0}^{E_1}\text{d}E\, \frac{\text{d}
    \Phi}{\text{d} E} A_\text{eff} (E) \label{mu_sig_dm}.
\end{equation}

Note that for simplicity, throughout this paper we neglect the effects of the
finite angular and energy resolution of CTA, as well as variations of the
effective area within the FoV.  As discussed above, the 68\% containment radius
of the point spread function (PSF) is around $\theta_{68\%}\simeq 0.3^\circ$ at
the lowest energies that we consider, and significantly better at high
energies.  This is significantly smaller than the RoIs that we adopt
(\textit{cf.}~Fig.~\ref{fig:rois}).  The energy resolution at 20 GeV is however
quite large ($\sigma(E)/E\simeq 50\%$), and should be taken into account when
interpreting the differential sensitivities that we quote below
(Sec.~\ref{sec:Results}).  We checked that smearing a $b\bar{b}$ spectrum with
a log-normal distribution with a width of $\sigma(\ln E)/\ln E = 50\%$
increases the peak signal flux relative to an assumed $E^{-2}$ background by a
factor of about 1.5; for $\sigma(\ln E)/\ln E = 20\%$ this drops to a factor of
1.1.  This is mostly due to event migration from smaller to higher energies.
Neglecting the finite energy resolution of CTA will hence affect our projected
limits at most by a few tens of percent.\footnote{For an analysis including the
effects of energy resolution see Ref.~\cite{Pierre14}.}

\section{Backgrounds}
\label{sec:Backgrounds}

The dominant backgrounds for DM searches at the GC are the GDE and, in case of
IACTs, the flux of cosmic rays that pass the photon cuts.  Here we discuss the
characteristics of these components, and describe how they enter our analysis.

\subsection{Cosmic-ray background}
\label{crbg}

CR electrons constitute an essentially irreducible background for gamma-ray
observations with IACTs, because the electromagnetic cascades that they induce
are practically indistinguishable from those caused by gamma rays.  A possible
way to discriminate between the two is to use the detection of Cherenkov light
from the primary particle as a veto.  This is however beyond the reach of
current and next-generation instruments~\cite{Consortium:2010bc}.  Due to their
relatively soft spectrum, CR electrons are especially relevant at low energies,
were they are responsible for most of the measured flux of particles that are
either gamma rays or electrons.  We parameterise the electron flux per unit
energy and solid angle as
\begin{equation}
    \frac{\text{d}^2\phi}{\text{d}E \text{d}\Omega} = 1.17 \times
    10^{-11}\left(\frac{E}{\text{TeV}}\right)^{-\Gamma} (\text{GeV\,cm}^2\,
    \text{s\,sr})^{-1}\;,
\end{equation}
with $\Gamma = 3.0$ for $E < 1 $\,TeV, and $\Gamma = 3.9$ for $E>1$\,TeV
\cite{Abdo:2009zk, Aharonian:2008aa}.

% Cosmic Rays and Extensive Air Showers, Stanev
% http://www.gae.ucm.es/~emma/docs/tesina/node17.html

Interactions between CR protons and the Earth's atmosphere initiate hadronic
air showers, which are generally very different from electromagnetic showers.
The initial hadronic interaction produces a large number of relatively slow
pions.  About 1/3 are neutral pions that decay into photon pairs and initiate
electromagnetic sub-showers, whereas the other 2/3 are charged pions, which
either decay and dump most of their energy into neutrinos, or interact again.
The resulting shower is much more widely distributed than those initiated by
gamma rays, allowing the shower shape to be used as a means for discriminating
between electromagnetic and hadronic showers.  Hadronic CRs are however far
more numerous than CR electrons and gamma rays.  The corresponding proton
rejection factor (\ie the fraction of incoming protons that are incorrectly
identified as photons/electrons) is of the order $\epsilon_p\sim 10^{-2}$ to
$10^{-1}$ for current instruments.  The electrons produced in electromagnetic
cascades from neutral pion decay are the dominant contributors to the
detectable Cherenkov light.  This means that if a cosmic-ray proton with a
given energy is mistakenly reconstructed as a gamma-ray, the gamma-ray is
usually reconstructed to have roughly a third as much energy as the actual
incoming proton.  

Here we adopt the proton flux per unit energy and angle
\citep{Hoerandel:2002yg}
\begin{equation}
    \frac{\diff\phi}{\diff E \diff\Omega} = 8.73 \times 10^{-9}
    \left(\frac{E}{\text{TeV}}\right)^{-2.71} (\text{GeV\,cm}^2\,\text{s\,sr})^{-1} \;,
\end{equation} 
which we shift to lower energies by a factor of 3 to account for the reduced
Cherenkov light emitted by hadronic showers \cite{Fegan:1997db}.  We note that
in principle heavier CR species are relevant, especially $^4$He, but these can
be effectively accounted for by increasing the proton cut efficiency factor
$\epsilon_p$.

Throughout the present analysis, we will adopt a proton cut efficiency of
$\epsilon_p=10^{-2}$ for definiteness. We note, however, that assuming that
the proton rejection factor is constant across all energies is a
simplification. At lower energies shower shape cuts are less effective at
discriminating CR protons from gamma-rays.  Thus our CR proton background is
underestimated at low energy, yielding a slight over-estimation of sensitivity
at low energies. However, even with this approximation we are still able to
reproduce the background rates shown in Ref.~\cite{Bernlohr:2012we} for Array I
to within a factor of two, for energies of up to 10\,TeV.  We note that
when switching off protons entirely (which corresponds to $\epsilon_p=0$) the
background rates remain practically unchanged at energies below 2\,TeV, as in our setup they
are dominated by electrons. We also checked that with our
assumptions we can reproduce the point source sensitivity for Array I to within
a factor of two below 100\,GeV, and within a few tens of percent above.

While the flux of CRs is isotropic across the sky, their acceptance by CTA
is not. The characterization of this variation in CR acceptance rates is a
non-trivial task, and methods to deal with non-isotropic acceptance rates are
discussed in \cite{Berge:2006ae}. For this analysis we make the simplification
of assuming an isotropic CR acceptance, though we note that our statistical
framework can accommodate anisotropic uncertainties in the combined CR and
gamma-ray acceptance, as discussed in Subsection \ref{subsec:Statistics}.

As formal gamma-ray and CR electron efficiencies we adopt
$\epsilon_\gamma=\epsilon_e=1$, and note that these values (along with
$\epsilon_p$) are defined with respect to the effective area of Array I from
Ref.~\cite{Bernlohr:2012we}.

\begin{figure}[t]
    \begin{center}
        \includegraphics[width=0.9\linewidth]{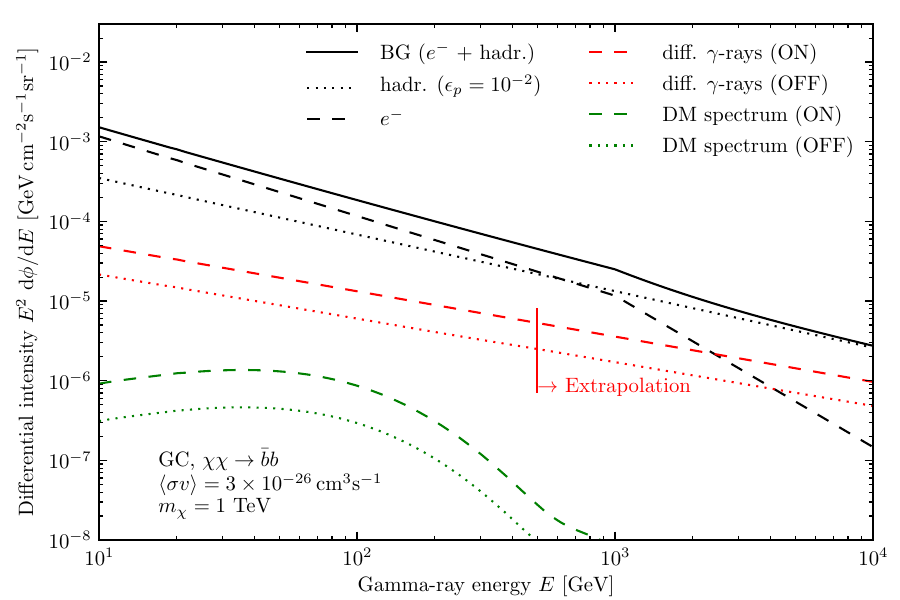}
    \end{center}
    \caption{Background fluxes relevant for our analysis. Isotropic CR
    backgrounds are shown in black: protons with an assumed cut efficiency of
    $\epsilon_p=10^{-2}$ (black dotted), electrons (black dashed), and total
    isotropic CR backgrounds (black solid). Galactic diffuse emission (GDE) is
    shown in red, and an example spectrum of DM annihilating to gamma-rays via
    $b\bar{b}$ is shown in green.  We give the DM and GDE curves for the ON and
    OFF regions defined in the Ring Method, as described in Section
    \ref{sec:DarkMatterSignals}.  Beyond 500\,GeV, we extrapolate the GDE
    spectrum using a simple power law.}
    \label{fig:backgrounds}
\end{figure}

In Fig.~\ref{fig:backgrounds}, we show the contributions from CR electrons and
protons assuming $\epsilon_p=10^{-2}$, and their sum.  Note that in the highest
energy range that we consider in this work, 6.7 to 10\,TeV, 100\,hr of CTA
observations lead to approximately $7.0\times10^3$ or $1.5\times10^{3}$ CR
background events in the OFF and ON regions, respectively.  For comparison we
also show the gamma-ray flux from a reference DM annihilation signal, based on
a thermal annihilation cross section and a DM mass of 1\,TeV, within the ON and
OFF regions employed in our version of the Ring method (\textit{cf.}~Section
\ref{sec:DarkMatterSignals}).  The CR foreground is stronger than the DM signal
by 2--3 orders of magnitude.  Fig.~\ref{fig:backgrounds} also shows a reference
GDE gamma-ray background spectrum, which we discuss next.

\subsection{Diffuse gamma-ray background}
\label{subsec:gde}

In 2006 HESS discovered diffuse gamma-ray emission from the GC at energies
of 0.2--20\,TeV \cite{Aharonian:2006au}.  The emission was found to be
correlated with molecular clouds in the central $200$\,pc of the Milky Way, and
is confined to Galactic latitudes $|b|<0.3^\circ$ and longitudes $|\ell| <
0.8^\circ$.  The spectrum suggests a hadronic origin.  The absence of evidence
for diffuse emission outside this window strongly influenced the choice of
search regions for DM signals in previous analyses~\cite{Abramowski:2011hc,
Doro:2012xx}.

Below 100\,GeV, the GDE has been measured extremely well by the
\textit{Fermi}-LAT \cite{FermiLAT:2012aa}.  At these energies, it is expected
to be dominated by $\pi^0$ decay from proton-proton interaction and
bremsstrahlung.  Diffuse gamma rays below 100\,GeV are an important background
in searches for TeV-scale DM, particularly with CTA, which will have an energy
threshold of tens of GeV.

To estimate the amount of GDE in different sky regions, and to study its impact
on DM searches at the GC, we adopt the P7V6 GDE model by the LAT team.  This
model extends up to 500\,GeV, above which we use a simple power-law
extrapolation.\footnote{This is not relevant to our discussion except at very
high DM masses, close to 10\,TeV. See
\url{http://fermi.gsfc.nasa.gov/ssc/data/access/lat/BackgroundModels.html} for
details on the BG model.}  The P7V6 GDE model was fit using data between 50
MeV and 50 GeV, and structures with extensions of less than $2^\circ$ were
filtered out. As much of our analysis is done outside the bounds of the
original data used to build the GDE model, e.g.~at energies up to 1 TeV, and
within $2^\circ$ of the GC, the effects of GDE on our analysis should be
considered as approximate only.  However we do not expect large changes in our
results when using more realistic GDE models: even factor two changes in the
background fluxes would affect our limits only at the $\sim30$--$40\%$ level.
We leave a more detailed study of prospects for a combined CTA +
\textit{Fermi}-LAT diffuse analysis for future work.

In Fig.~\ref{fig:backgrounds}, we show the contribution that we inferred from
this particular GDE model in the ON and OFF regions.  For 100\,hr of CTA
observations, at energies of 6.7--10\,TeV this corresponds to about
$1.2\times10^3$ and $4.9\times10^{2}$ events respectively in the OFF and ON
regions.  This is a factor of ten higher than the reference DM signal at its
peak value, and larger in the ON than in the OFF region.  For these reasons,
the GDE is a very important background that should not be neglected in DM
searches at the GC.

\section{Analysis}
\label{sec:Analysis}

\subsection{Analysis regions and $J$ factors}
\label{subsec:RegionsAndJfactors}

For our version of the Ring method, we begin with the standard annulus of Ref.\
\cite{Doro:2012xx}, with an inner radius $r_1$ and outer radius $r_2$.  The
centre of this annulus is offset from the GC ($b_0 = \ell_0 = 0^\circ$) by some
Galactic latitude $b$.  We then consider a circular region centred on the GC,
with some radius $\Delta_\text{cut}$.  The area in which the annulus and this
circular region intersect is what we refer to as the `ON' region.  The `OFF'
region consists of the remaining part of the annulus, outside the central disc.
We adopt the parameters optimised for Array E in Ref.\ \cite{Doro:2012xx}:
$b=1.42^\circ$, $r_1 = 0.55^\circ$, $r_2 = 2.88^\circ$ and $\Delta_\text{cut} =
1.36^\circ$.  Further, we exclude the Galactic disc within $|b|\leq0.3^\circ$
from both the ON and OFF regions, as per Ref.\ \cite{Doro:2012xx}.  The
resulting two RoIs can be seen in the left panel of Fig.~\ref{fig:rois}.  The
corresponding solid angles and $J$ factors are $\Delta\Omega_\text{ON} = 1.2
\times 10^{-3}$\,sr, $\Delta\Omega_\text{OFF} = 5.6 \times 10^{-3}$\,sr,
$J_\text{ON} = 7.4 \times 10^{21}\,\text{GeV}^2\,\text{cm}^{-5}$ and
$J_\text{OFF} = 1.2 \times 10^{22}\,\text{GeV}^2\,\text{cm}^{-5}$.

For our morphological analysis, we take the area covered by these two RoIs, and
divide it into $1^\circ \times 1^\circ$ squares.  We horizontally merge the
various leftover regions resulting from this dissection into adjacent regions,
yielding a total of 28 RoIs.  These spatial bins are shown in the right panel
of Fig.~\ref{fig:rois}.

For comparison, we also consider DM annihilation in the Draco dwarf spheroidal
galaxy, which, to a good approximation, is a point source to both CTA and
\textit{Fermi} at low energies (in the upper parts of their respective energy
ranges, both would observe Draco as a somewhat extended source).  For this
analysis, we use the $J$-factor and solid angle from Table 1 of Ref.\
\cite{Ackermann:2013yva}: $\Delta\Omega_\text{Draco} = 2.4 \times 10^{-4}$\,sr,
$J_\text{Draco} = 6.31 \times 10^{18}\text{\,GeV}^2\,\text{cm}^{-5}$.

\subsection{Statistical framework}
\label{subsec:Statistics}  

We use a binned Poisson likelihood function for comparing a DM model $\model$
to (mock) data $\data$
\begin{align}
    \mathcal{L}\left(\model|\data\right) = \prod_{i,j} \frac{\modelij^\dataij}{\dataij!} \exp(-\modelij). 
    \label{eq:basicLpoisson}
\end{align}
Here the predictions of model $\model$ are the number of events $\modelij$ in
the $i$th energy bin and the $j$th RoI, which are compared to the corresponding
observed counts $\dataij$.  We use 15 logarithmically-spaced energy bins,
extending from 25\,GeV to 10\,TeV.  Depending on the analysis (Ring or
morphological), we use either two (Ring) or 28 (morphological) spatial bins
(\ie RoIs).

Each model prediction is composed of 3 parts: a gamma-ray signal resulting from
DM annihilation (Eq.~\ref{mu_sig_dm}), an isotropic cosmic-ray background, and
the GDE. In our statistical analysis each of these components can be rescaled
via a parameter: $\langle \sigma v \rangle$ for the DM gamma-ray signal, and
linear rescaling factors $R_\text{CR}$ and $R_\text{GDE}$ for the isotropic
cosmic-ray background and the GDE respectively, 
\begin{equation}
    \modelij = \modelij^\text{DM} + \modelij^\text{CR}R_{\text{CR},i} +
    \modelij^\text{GDE}R_{\text{GDE},i}.
\end{equation}
Given that CR flux is isotropic to a good approximation and that we adopt
the simplifying assumption that acceptance are isotropic, the
$\mu_{ij}^\text{CR}$s are trivially related by
$\mu_{ij}^\text{CR}/\mu^\text{CR}_{ik} = \Omega_{j}/\Omega_{k}$, where
$\Omega_j$ denotes the angular size of the $j$th RoI. Note that we do not vary
the relative normalisations of the CR electron and proton spectra, only their
sum.  We also do not allow the GDE to be rescaled independently in each RoI, as
this would simply allow the GDE to adjust to the data in its entirety in every
bin, effectively non-parametrically.

Systematic uncertainties in the signal rates can be accounted for by
multiplying the predicted signals $\modelij$ by scaling parameters $\alphaij$
and $\betai$, then profiling the likelihood over their values.  We assume
Gaussian nuisance likelihoods for all $\alpha$ and $\beta$, with respective
variances $\sigma_\alpha^2$ and $\sigma_\beta^2$ independent of $i$ and $j$.
Strictly, the distributions should be log-normal so that they go smoothly to
zero as $\alpha$ and $\beta$ go to zero, but for small $\sigma$ this makes
practically no difference; typical values of $\sigma$ that we will consider in
the following section are $\le 0.03$, well within the range where this is a
good approximation.  With these scaling parameters, the likelihood function
becomes
\begin{equation}
    \mathcal{L}\left(\model,\boldsymbol{\alpha},\boldsymbol{\beta}|\data\right)
    = \prod_i 
    \frac{1}{\sqrt{2\pi}\sigma_\beta}
    e^{- \frac{(1-\betai)^2}{2 \sigma_\beta^2}}
    \prod_j
    \frac{\left(\modelij\alphaij\betai\right)^\dataij}{\sqrt{2\pi}\sigma_\alpha\cdot\dataij!}
    e^{-\modelij\alphaij\betai}\,e^{-\frac{(1-\alphaij)^2}{2 \sigma_\alpha^2}}\;.
    \label{eq:Likelihood_wSyst}
\end{equation}
This formulation accounts for systematic uncertainty on any factor that enters
linearly in the calculation of the total signal, such as effective area.  The
parameters $\alphaij$, which vary across both energy bins and RoIs, account for
e.g.~uncertainties related to non-uniformities in the acceptance of CTA within
a FoV.  We will refer to all uncertainties described by $\alphaij$ as
\emph{differential acceptance uncertainties}.  Reasonable values for
$\sigma_\alpha$ are of the order of a few percent~\cite{Funk:2012ca,
Dickinson:2012wp}.  The parameters $\betai$, on the other hand, describes
systematic uncertainties for a given energy that apply equally over the whole
FoV.

Here we are most interested in the cross-RoI systematics, as systematics that
apply equally across the whole FoV will essentially just degrade an entire
$n\sigma$ confidence-level sensitivity curve by an offset of less than
$n\sigma_\beta$ (although the rescaling could differ with energy, if
$\sigma_\beta$ were permitted to vary with energy).  We therefore investigate
the impacts of allowing each $\alphaij$ to vary independently, and simply set
$\betai=1$ for all $i$.  The impacts of e.g.~an energy-dependent systematic
uncertainty on the \textit{Fermi} or CTA effective areas could be easily
accounted for by also profiling over each $\betai$.

Perhaps the largest source of uncertainties is the modelling of the CR
acceptance. While in this analysis we assume an isotropic CR acceptance, this
is a simplification, as noted above. This anisotropic acceptance could be
incorporated in our analysis framework, but a detailed discussion is beyond the
scope of this work.  

Taking the likelihood function Eq.\ \eqref{eq:Likelihood_wSyst}, we perform a
maximum likelihood analysis across $\alphaij$, $\langle \sigma v \rangle$,
$R_\text{CR}$, and in certain cases, $R_\text{GDE}$. We can determine the
maximum likelihood value of $\alphaij$ simply by solving
\begin{equation}
    \frac{\diff \mathcal{L}\left(\modelij, \alphaij, \betai | \dataij
    \right)}{\diff \alphaij} = 0,
\end{equation}
and demanding that $\alphaij$ be positive. This gives
\begin{align}
    \label{alphaval}
    \alphaij = \frac{1}{2} \left(1 - \sigma_\alpha^2 \modelij\betai + \sqrt{1 +
    4 \sigma_\alpha^2 \dataij  - 2\sigma_\alpha^2 \modelij\betai  +
    \sigma_\alpha^4\modelij^2\betai^2} \right).
\end{align}

In our morphological analysis we divide the FoV into bins of $1^\circ$
squared, as laid out in Subsection \ref{subsec:RegionsAndJfactors} above.  The
systematic acceptance uncertainty in each bin is described by independent
normal distributions in the likelihood function, which implies that the
correlation scale of these uncertainties is of the order of $ \sim1^\circ$.
Since the uncertainties are treated as statistically independent in the
individual bins, they tend to average out.  In fact, decreasing the solid angle
of the bins by a factor of four would be roughly equivalent to reducing the
differential acceptance uncertainty by a factor of two.  Hence, when we quote
these uncertainties for our morphological analysis, it is important to keep in
mind that they refer to $\sim1^\circ$ correlation scales, which we adopt here
as reference value.  A full exploration of the effect of different correlation
scales would require more detailed knowledge of the detector performance, and
is beyond the scope of this analysis. 

The mock data $\data$ that we use for deriving sensitivities in the following
section assume a fixed isotropic cosmic-ray background component with
$R_{\text{CR},i} = 1$ in all bins, and no contribution from DM annihilation.
Depending on the analysis, we either include no GDE ($R_{\text{GDE},i} = 0$ for
all $i$) in the mock data, or a fixed GDE with $R_{\text{GDE},i} = 1$.  The
mock data sets that we employ are not Poisson realisations of a model with
these assumptions, but rather simply the expectation number of events given
these assumptions; this has been referred to as the ``Asimov data set''
\cite{Cowan11}.

We calculate $95\%$ confidence level (CL) upper limits by increasing the signal
flux (or annihilation cross-section) from its best-fit value, whilst profiling
over the remaining parameters, until $-2\ln\mathcal{L}$ changes by 2.71.  In
the case of the differential point-source sensitivity that we discuss in the
next section, for a $95\%$ CL upper limit we also require at least one energy
bin to contain at least 3.09 events~\cite{Feldman:1997qc}.  In this
calculation, we determine background rates over the $80\%$ containment region
of the PSF, following e.g.~Ref.~\cite{Bernlohr:2012we}.  Note that we neglect
instrumental systematics when evalulating \textit{Fermi}-LAT sensitivities
(setting $\alphaij=\betai=1$).

\subsection{Background treatment}
\label{subsec:DiffEmAnalysis}

In both our Ring and morphological analyses, we allow the isotropic cosmic-ray
background rescaling factor $R_{\text{CR},i}$ to vary between 0.5 and 1.5 in
our fits.  We then profile the likelihood over this parameter in each energy
bin.

Including the GDE in our analysis is more complicated, as in principle, the
data-driven GDE model derived by \textit{Fermi} could already contain some
contribution from DM annihilation.  Therefore, to gauge the impact of the
diffuse emission upon current analysis methods, for our analysis with the Ring
method we inject the GDE into our mock dataset $\data$, with $R_{\text{GDE},i}
= 1$ in all bins.  We then carry out a full analysis with a model $\model$
where the GDE normalisation is left free to vary, \ie $R_{\text{GDE},i}$ is
left free in the fits, but we require it to be non-negative.  The idea in this
exercise is to leave the analysis method as much as possible unaltered relative
to previous analyses, but to make the mock data fed into the method more
realistic.  Leaving the GDE free to vary in each energy bin produces the most
conservative constraints and avoids assumptions on the GDE energy spectrum.
The results of this analysis are given in Sec.~\ref{subsec:GCSensOnOff}. 

Our morphological analysis allows a more refined inclusion of GDE.  Using more
than two RoIs allows us to better exploit the shape differences between the
GDE, which is concentrated along the Galactic plane, and the DM annihilation
signal, which is spherically distributed around the GC. For this analysis we
again include the GDE in the mock data with $R_\text{GDE} = 1$ in all bins, and
allow the individual $R_{\text{GDE},i}$ values to vary in each energy bin.  To
implement this analysis we use lookup tables: for each DM mass, and each energy
bin and RoI, we consider a range of DM cross-sections.  For each cross-section
value, we calculate the maximum likelihood point when profiling numerically
over $R_\text{BG}$ and $R_\text{GDE}$, and analytically over all $\alphaij$
using Eq.\ \eqref{alphaval}.  Storing these values then gives us a partial
likelihood as a function of cross-section in that bin, for each DM mass.  For a
given DM mass, we can then combine the partial likelihoods in different bins to
determine cross-section limits at arbitrary confidence levels.  The results of
this analysis are given in Sec.\ \ref{subsec:GCSensMorpho} and
\ref{subsec:GCsigmav}.

\section{Results}
\label{sec:Results}

To keep the discussion as independent from specific DM scenarios as possible,
in this section we will often quote the \emph{differential sensitivity}, which
is the sensitivity to a signal in an isolated energy bin.  This measure is
commonly used in the gamma-ray community (see e.g.~Ref.~\cite{Funk:2012ca}).
Here we consider energy bins with a width of $\simeq0.17$ dex (approximately
six bins per decade), and quote sensitivities in terms of projected one-sided
$95\%$ CL upper limits.

We will start with a discussion of the point source sensitivity, which is
relevant for DM searches in dwarf spheroidal galaxies.  The remaining part of
this section will then discuss DM searches at the GC.  Finally, we will present
projected upper limits on the DM annihilation cross-section for various
annihilation channels and DM profiles.

\subsection{Point source sensitivity}
\label{subsec:PointSourceSens}

\begin{figure}
    \begin{center}
        \includegraphics[width=0.9\linewidth]{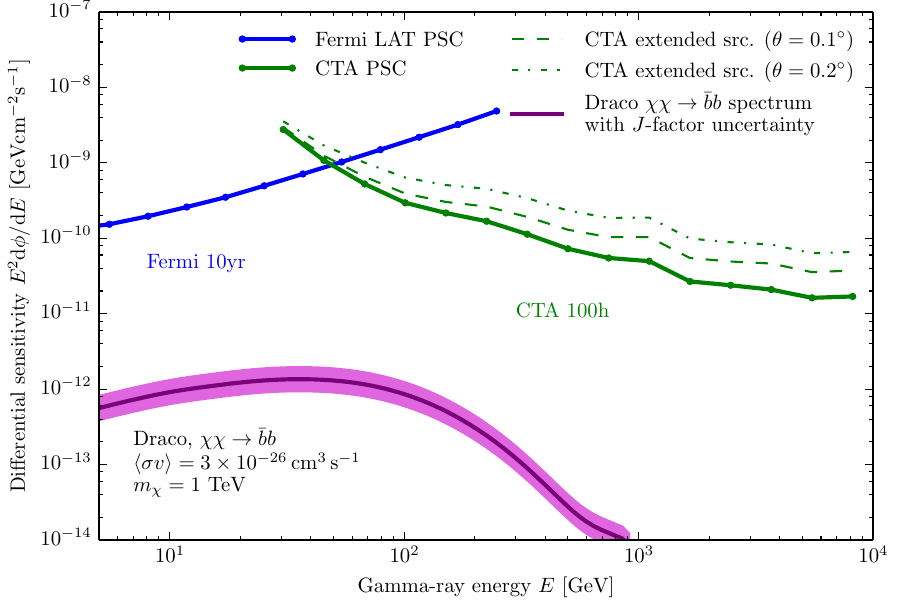}
    \end{center}
    \caption{Differential sensitivity of \textit{Fermi}-LAT (blue) and CTA
    (green) point source observations, in terms of 95\% CL upper limits.  The
    \textit{Fermi}-LAT differential sensitivity assumes an observation time of
    10 years, with energy bins of $\Delta\log_{10}E = 0.165$ size between
    $E=5.0$\,GeV and 300\,GeV, and includes only statistical errors.  The CTA
    sensitivities assume an observation time of 100 hours, include 1\%
    differential acceptance uncertainty, with bins of $\Delta\log_{10}E=0.173$
    size between $E=25$\,GeV and $10$\,TeV.  Sensitivities for extended sources
    of different sizes are also shown for CTA (green dashed and dot-dashed).
    The DM annihilation spectrum from the Draco dwarf galaxy uses a $J$-factor
    of $6.31^{+3.24}_{-1.94} \times 10^{18}\, \text{GeV}^2 \text{cm}^{-5}$ from
    Table 1 of \cite{Ackermann:2013yva}, with the band covering the uncertainty
    range.}
    \label{fig:diff_sens_PSC}
\end{figure}

Some of the most powerful targets for indirect DM searches with gamma rays are
dwarf spheroidal galaxies.  In order to compare the potential of CTA with the
abilities of current instruments like \Fermi-LAT, it is instructive to consider
their differential point source sensitivity.  In Fig.~\ref{fig:diff_sens_PSC},
we show the differential sensitivity of CTA, assuming an observation time of
100 hours, compared to the one of \Fermi-LAT after ten years of observation
(assuming 20\% of the time is spend on the source). Due to its much larger
effective area, CTA will outperform \Fermi-LAT at energies above about
$100\rm\,GeV$, where \Fermi\ becomes limited by photon statistics.  

For comparison, we show the signal flux expected from the dwarf spheroidal
galaxy Draco, in the case of a DM particle with $m_\chi=1\rm\,TeV$ mass,
annihilating into $b\bar{b}$ final states with an annihilation cross section of
$\langle\sigma v\rangle = 3\times10^{-26}\rm\,cm^3\,s^{-1}$. Draco is one of
the most promising targets, and the envelope shows the uncertainty in the
projected signal flux, which is primarily related to its overall mass (taken
from Ref.~\cite{Ackermann:2013yva}).

Dwarf spheroidal galaxies have half-light radii of a few times $0.1^\circ$ (see
e.g.~Ref.~\cite{Cholis:2012am}), and at high energies will appear as extended
sources for CTA.  For comparison, we show the impact on the expected
sensitivity in Fig.~\ref{fig:diff_sens_PSC}.  We derive these sensitivity
curves by assuming that a dwarf is observed by CTA to have an angular extent
given by the sum in quadrature of the 68\% PSF containment radius and the
intrinsic angular extent indicated in the figures. This effect will degrade the
sensitivity by a factor of a few at TeV energies.  As can be seen in
Fig.~\ref{fig:diff_sens_PSC}, for a gamma-ray spectrum from hadronic processes
like annihilation into $b\bar{b}$, CTA will outperform \Fermi\ for DM masses
above about $1\rm\,TeV$.  However, the sensitivity of CTA will still fall about
two orders of magnitude short of testing the canonical thermal cross-section.  

In order to understand how the limits scale with observing time, it is helpful
to realise that in this transition regime the limiting factor for CTA is the
enormous background of CR electrons (and to a lesser degree unrejected protons
and light nuclei; see Fig.~\ref{fig:backgrounds}). These CR electrons can
easily swamp weak sources even if the number of events that are detected from
the source is much larger than in case of \Fermi-LAT.  Indeed, apart from the
difference in the effective area, the main difference between space-based and
ground-based gamma-ray telescopes is their respective abilities to reject
backgrounds.  This is relatively simply realised by an anti-coincidence
detector in the case of \Fermi, but extremely challenging in the case of
Cherenkov Telescopes like CTA.  This means that even a larger observing time
with CTA would not significantly affect the point-source sensitivity at the low
energies relevant for TeV DM searches.  In contrast, much longer observation
times with space-based and practically background-free instruments like \Fermi\
could (at least in principle) improve on existing limits by orders of
magnitude.

\subsection{Galactic centre sensitivity with Ring method}
\label{subsec:GCSensOnOff}

The most intense signal from DM annihilation is expected to come from the GC.
The large CR background will in that case be of lesser relevance than for
observations of dwarf spheroidal galaxies, making the GC a particularly
promising target for CTA.  Previous analyses found that CTA will improve
existing limits (with the strongest ones coming currently from HESS) by an
order of magnitude or so~\cite{Doro:2012xx, Wood:2013taa, Pierre14}.  However,
all existing analyses have neglected the impact of the GDE, which is strongest
in the direction of the Galactic center.  We will demonstrate here that this in
fact has a significant impact on DM searches with CTA.

\begin{figure}
    \begin{center}
        \includegraphics[width=0.9\linewidth]{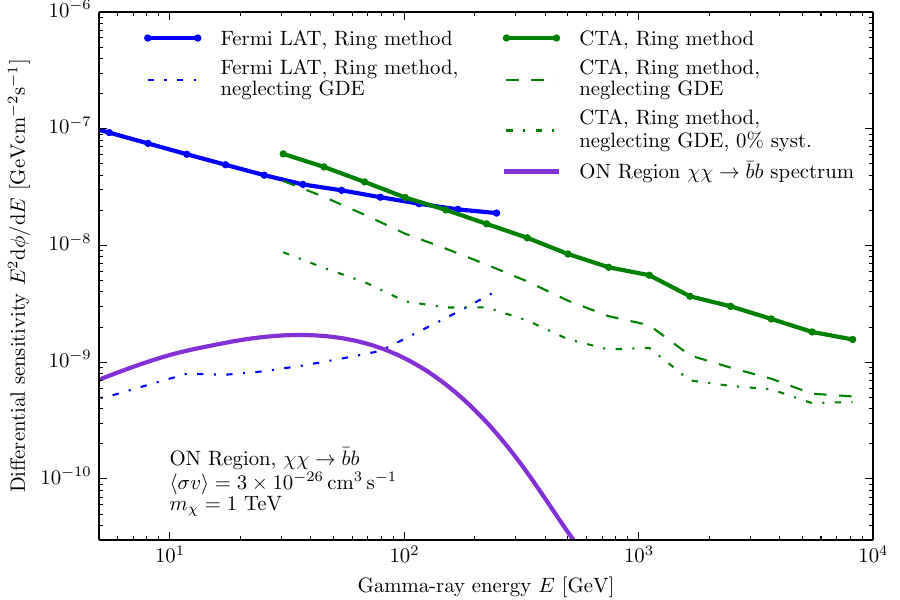}
    \end{center}
    \caption{Differential sensitivity of \textit{Fermi}-LAT (blue) and CTA
    (green) for GC observations, using the Ring method as defined in Section
    \ref{sec:DarkMatterSignals}. The two upper solid lines show our baseline
    estimate for the sensitivity, with galactic diffuse emission and a
    differential acceptance uncertainties of 1\% included. For comparison, the
    dashed green line shows the differential sensitivity for CTA neglecting the
    GDE, but still including systematics of 1\%; the dash-dotted lines in blue
    and green show the differential sensitivity for \textit{Fermi}-LAT and CTA
    respectively, neglecting GDE and instrumental systematics. Observation
    times and energy binning are the same as in Fig.~\ref{fig:diff_sens_PSC}.
    The purple line shows the gamma-ray spectrum from the ON region assuming a
    DM mass of 1\,TeV and an annihilation cross section of $\langle \sigma v
    \rangle = 3 \times 10^{-26} \text{cm}^3 \text{s}^{-1}$ to $b\bar b$, using
    the Einasto profile from Eq.~\eqref{einasto}.}
    \label{fig:diff_sens_GC_OnOff_Fermi}
\end{figure}

In Fig.~\ref{fig:diff_sens_GC_OnOff_Fermi}, we show the differential
sensitivity for a DM signal in the GC that we obtain when using our version of
the Ring method (\cite{Doro:2012xx}, \textit{cf.}~Fig.~\ref{fig:rois}), under
various assumptions about the GDE and instrumental systematics.  When creating
mock data for our baseline analysis, we include individual estimates for the
GDE in the different RoIs.  The Ring method is only sensitive to integrated
fluxes measured in the ON and OFF regions.  In our likelihood fit to the mock
data the normalisations of the DM and the GDE components are therefore
degenerate, as these two components in general contribute a larger flux to the
ON than the OFF region -- whereas variations in the average intensity in both
regions can be absorbed by slight changes in the CR background normalisation.
An increased DM contribution can be compensated for by a smaller GDE
contribution, and vice versa.  This degeneracy breaks when the GDE contribution
drops to zero, at which point the $-2\ln \mathcal{L}$ increases with increasing
DM annihilation cross-section, which then leads to a conservative upper limit
on the DM flux.

The curves that we show in Fig.~\ref{fig:diff_sens_GC_OnOff_Fermi} where the
GDE has not been included are based on neglecting GDE in the the mock data and
the subsequent analysis (setting $R_\text{GDE}=0$ everywhere).  We see here
that in the case of the Ring method, neglecting the GDE artificially improves
the projected sensitivity by a factor of up to about 3.  Furthermore, neglecting
instrumental systematics (as in Ref.~\cite{Wood:2013taa}) increases the
sensitivity again by another factor of a few.  For comparison, we again show
the flux expected from a DM particle with $1\rm\,TeV$ mass, annihilating into
$b\bar{b}$ at the canonical thermal rate.

Since the GDE component is varying independently in each energy bin, our
statistical framework is insensitive to spectral information that could help to
discriminate between the GDE and the DM signal. Including this information
could potentially increase the sensitivity of the Ring method, but would
require precise assumptions on the poorly known spectrum of the GDE. By
including spectral information, the authors of Ref.~\cite{Aleksic:2013xea} were
able to improve the limits derived from observations of Segue 1 by MAGIC by a
factor of between 1.9 and 3.3, compared to a standard analysis using only
spatial information.  Applying similar analysis techniques to CTA could yield
in the best case a comparable improvement in sensitivity.  However, the
uncertainty in the GDE spectra has to be carefully addressed in that case,
which we leave to future work.  

\subsection{Galactic centre sensitivity with multi-bin morphological method}
\label{subsec:GCSensMorpho}

It is instructive to see the results that one would obtain by applying the Ring
method to \Fermi-LAT instead of CTA data.  As shown in
Fig.~\ref{fig:diff_sens_GC_OnOff_Fermi}, the resulting sensitivity curve simply
continues the curve from CTA to lower energies.  This can be readily
understood, as in both cases the actual limits are driven by the same GDE.  We
can see that the GDE is the factor holding back the Ring analysis. This
motivates us to consider a procedure capable of taking into account
morphological differences between the GDE and DM signal, which is what we
discuss next.\footnote{Considering the energy spectrum of the signal as we do
here of course also improves CTA limits \cite{Pierre14}; all our analyses are
carried out including this information.  Once a signal is detected however, if
gamma-ray lines or virtual internal Bremsstrahlung seem relevant, a primarily
spectral analysis would be preferable~\cite{Bergstrom:2012vd,
Abramowski:2013ax}.}

\begin{figure}
    \begin{center}
        \includegraphics[width=0.9\linewidth]{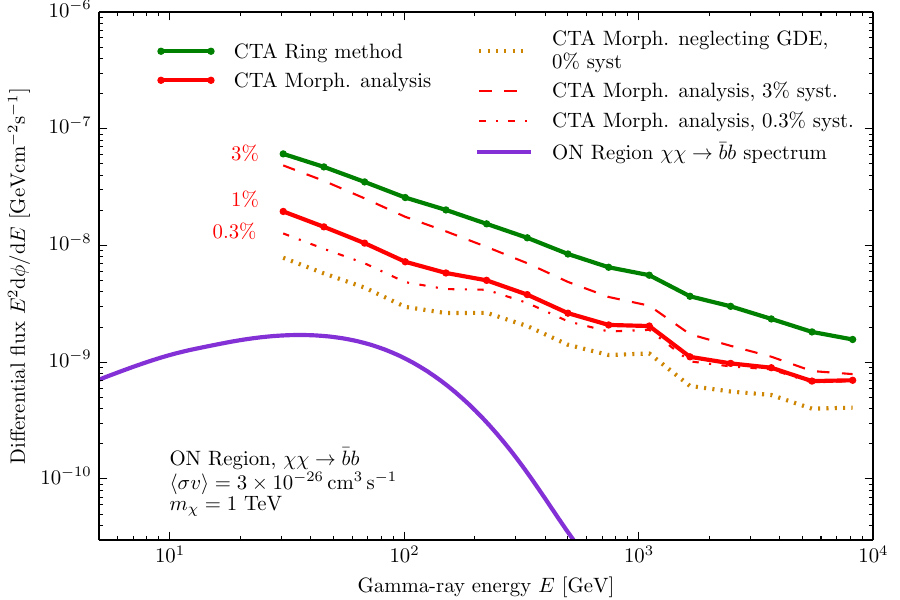}
    \end{center}
    \caption{Same as Fig.~\ref{fig:diff_sens_GC_OnOff_Fermi}, but comparing our
    previous Ring analysis (green) with our morphological analysis (red), again
    assuming 100\,hr of observation. The solid lines show our baseline estimate
    for the sensitivity of the two analysis methods, with galactic diffuse
    emission included and differential acceptance uncertainties of 1\%. The
    dotted orange line shows the same morphological analysis as our baseline
    estimate but neglecting the GDE and assuming 0\% systematics. Also shown are
    sensitivities produced using the morphological analysis including GDE but
    using systematics of 3\% (dashed) and 0.3\% (dash-dotted) instead.}
    \label{fig:diff_sens_GC_Morpho}
\end{figure}

Here we take an important first step towards an improved DM search strategy for
CTA by proposing a morphological analysis of the gamma rays from the GC.  We
will estimate the \emph{most optimistic} limits that one can obtain on DM
annihilation in the presence of the known GDE, assuming that the morphology of
the GDE is perfectly understood.  To this end, we define the 28 subregions
distributed as shown in the right panel of Fig.~\ref{fig:rois}.  They are all
part of the original RoI from the Ring method. This has the advantage that the
required observation strategy and results are directly comparable.

Fig.~\ref{fig:diff_sens_GC_Morpho} shows that with our morphological analysis,
the projected differential sensitivity is better by a factor of three than what %
can be obtained with the Ring method.  Note that the limiting factor in our
results, at least for sub-TeV energies, is now not the GDE but differential
acceptance uncertainties, namely relative systematic uncertainties in the
photon acceptance in different regions of the \emph{same} FoV.  For this
uncertainty, we assume $1\%$ throughout, which is a rather realistic value.
This might however vary by a factor of a few up or down, depending on the
experimental details (\textit{cf.}~Sec.\ \ref{subsec:Statistics}).  Indeed,
varying the systematic uncertainty in a reasonable range has a significant
impact on the actual projected constraints. Note that systematics of 0.3\%
give results extremely close to that of 0\% systematics. Also shown for
comparison in Fig.~\ref{fig:diff_sens_GC_Morpho} is our morphological analysis
assuming 0\% systematics and neglecting the GDE, i.e.~$\alpha_i = 1$ for all
$i$, and $R_{\text{GDE},i} = 0$ for all $i$ in both mock data and
model.\footnote{Although we introduced the morphological analysis method
primarily to improve limits in the presence of the GDE, we also compared its
performance to that of the Ring method in the case of no GDE and 0\%
systematics. In this case the morphological analysis produces limits that are
marginally better than those of the Ring method. This is expected, as the
smaller RoIs still provide an additional constraint on the spatial distribution
of the signal.}

\subsection{Projected cross-section limits}
\label{subsec:GCsigmav}

We now present our results in terms of limits on DM annihilation, in the common
$\langle \sigma v\rangle$-vs-mass plane, assuming DM annihilation into
different final states with a branching ratio of $100\%$.  First we provide
some context by summarising the most relevant previous work, and later
compare these to our own results. 

\begin{figure}
    \begin{center}
        \includegraphics[width=0.9\linewidth]{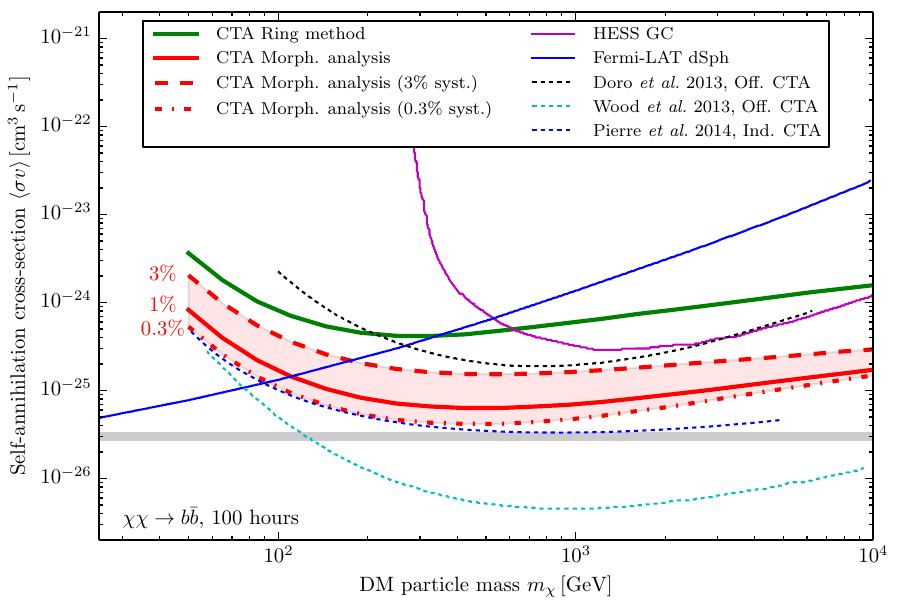}
    \end{center}
    \caption{Upper limits on the DM annihilation cross section using the
    previous Ring method (green) and our morphological analysis (red), assuming
    100\,hr of observation of the GC. The thick solid green and red lines are
    our baseline estimates of the limits attainable using the two analysis
    methods, assuming differential acceptance uncertainties of 1\% and
    including GDE. The red dashed and dot-dashed lines show the limits produced
    in the morphological analysis assuming 3\% and 0.3\% systematics,
    respectively. Also shown are current limits on the DM annihilation cross
    section (thin solid lines; \textit{Fermi}-LAT dwarf analysis in blue
    \cite{Ackermann:2013yva}, HESS GC observations in pink
    \cite{Abramowski:2011hc}), as well as various projected CTA limits,
    both official (thin dotted lines; Doro \textit{et al.} 2013 in
    black~\cite{Doro:2012xx}, Wood \textit{et al.} 2013 in cyan
    \cite{Wood:2013taa}), and independent (Pierre \textit{et al.}~2014,
    thin dotted line in dark blue \cite{Pierre14}).  For the sake of
    comparison, the CTA projections are rescaled to 100\,hr observation time
    and our adopted Einasto profile.}
    \label{fig:sv_limits_GC_OnOff_Morpho}
\end{figure}

In Fig.~\ref{fig:sv_limits_GC_OnOff_Morpho}, we show existing experimental
limits from the \textit{Fermi}-LAT satellite \cite{Ackermann:2013yva} and
HESS \cite{Abramowski:2011hc}, on DM annihilation into $b\bar{b}$.  In this
figure, all limits from the GC are rescaled to our baseline Einasto DM profile.
Projected limits correspond to 100\,hr observation time for CTA.  The
\Fermi-LAT limits reach the thermal cross section for DM masses below about 10
GeV.  The HESS limits are strongest close to 1 TeV, where they reach
$3\times10^{25}\rm\,cm^3\,s^{-1}$. 

Also shown are the projected limits for CTA from
Refs.~\cite{Doro:2012xx,Wood:2013taa, Pierre14}.  The analysis given in
Ref.~\cite{Wood:2013taa} assumes an observation time of 500\,hr, and we have
rescaled their limits to account for our baseline DM profile and observation
time of 100\,hr.  In fact, even after this rescaling, the projected limits from
Ref.~\cite{Wood:2013taa} remain the most optimistic; they apparently do not
account for systematic uncertainties or the effects of the GDE, and make use of
an extended array with 61 mid-sized telescopes. This extended array is assumed
to contain 36 extra mid-sized telescopes, as an additional, rather speculative,
US contribution on top of the baseline array. 

The limits presented in Ref.~\cite{Doro:2012xx} include no GDE and no spectral
analysis.  They were built upon a profile derived from the Aquarius simulation,
and therefore include an effective substructure boost compared to a regular NFW
profile.  We have removed this boost in order to allow direct comparison with
our results, and those of others.  The inclusion of substructure increases the
$J$-factor and thus also the signal, which results in a stronger limit.  As we
use identical regions of interest to those of \cite{Doro:2012xx}, we can
estimate the substructure boost factor by comparing ON region $J$-factors:
\cite{Doro:2012xx} give their result as $J_\text{ON,Aq.} = 4.68 \times 10^{22}
\text{GeV}^2 \text{cm}^{-5}$, while our smooth Einasto profile
(Sec.~\ref{sec:DarkMatterSignals}) gives $J_\text{ON, Ein.} = 7.41 \times
10^{21} \text{GeV}^2\, \text{cm}^{-5}$.  This yields a boost factor of 6.31; we
hence multiply the limits of Ref.~\cite{Doro:2012xx} by this factor for
presentation in Fig.~\ref{fig:sv_limits_GC_OnOff_Morpho}.

The limits presented in Ref.~\cite{Pierre14} are derived in a similar fashion
to the Ring method ones in the present analysis, including spectral analysis,
but neglect contributions from the GDE, systematics and proton CRs.

Our projected CTA upper limits on the annihilation cross-section using our
version of the Ring method are shown in
Fig.~\ref{fig:sv_limits_GC_OnOff_Morpho} by the thick green line.  In contrast
to Ref.~\cite{Doro:2012xx}, we \emph{include} the expected GDE as discussed
above (\textit{cf.}~Fig.~\ref{fig:diff_sens_GC_Morpho}). As a consequence, we
find somewhat weaker limits at intermediate masses than in this previous work.
From Fig.~\ref{fig:diff_sens_GC_OnOff_Fermi} one can see that neglecting  
GDE in the Ring method (in the mock data) falsely improves the sensitivity by a
factor of $\sim 2 - 3$, which agrees with the factor $\sim 3$ difference %
between the results of Ref.~\cite{Doro:2012xx} and our Ring method limit. The
different shape of the limits as function of mass is due to our lower energy
threshold, and the fact that we carry out a spectral analysis, whereas
Ref.~\cite{Doro:2012xx} used only one large energy bin.  This consequence of
including spectral information was noted previously in Ref.\ \cite{Pierre14},
whose limits have a very similar shape to our own. 

The projected CTA upper limit  produced by our morphological analysis is
shown in Fig.~\ref{fig:sv_limits_GC_OnOff_Morpho} by the thick red line (while
not shown on the figure, the limit neglecting GDE is a factor of $\sim 1.5$
below this line). It is perhaps surprising that our limit including GDE is
significantly better than that of Ref.~\cite{Doro:2012xx}, which neglects GDE
both in the mock data as well as in the analysis.   One reason is that using
more bins allows for better shape discrimination of the signal over the
isotropic background.  Another reason might be differences in the adopted
acceptance uncertainties (which are not quoted in Ref.~\cite{Doro:2012xx}).
However, the fact that our morphological analysis is an order of magnitude
weaker than the scaled limits of Ref.~\cite{Wood:2013taa} is primarily due to
the unrealistic array considered in that paper and their complete neglect of systematics.
\medskip

Interestingly, our projected constraints are actually also weaker than the
existing limits from HESS, which are based on 112 hours of GC observations.
This difference is likely due to two things.  The first is simply that the RoIs
adopted in the Ring method and in the HESS analysis are rather different.  The
second is that the HESS analysis is a true ON-OFF analysis, which neglects the
GDE by definition; this was only possible to do in a valid way in the HESS
analysis due to the instrument's high energy threshold and the fact that at
such energies, the GDE intensity observed by \Fermi-LAT happens (by chance?) to
be very similar in the rather complicated ON and OFF regions chosen by HESS.

Most importantly, moving from the Ring method to our morphological analysis
yields \emph{a sensitivity improvement by up to an order of magnitude}.  We
show in Fig.~\ref{fig:sv_limits_GC_OnOff_Morpho} that morphological analysis
improves the projected limits by up to a factor of ten for high DM masses
compared to the Ring method.\footnote{This is not directly apparent from
Fig.~\ref{fig:diff_sens_GC_Morpho}, where the difference is merely a factor of
three.  The reason is that a DM signal would appear in several energy bins
simultaneously, which strengthens the limits in the case where the GDE is
correctly modelled.}  This is mostly due to the fact that the large number of
subregions allows an efficient discrimination between the morphology of the GDE
and a putative DM signal.  Note that the projected limits again depend
critically on instrumental systematics, and as indicated by the band in
Fig.~\ref{fig:sv_limits_GC_OnOff_Morpho}.  This is due to the large number of
measured events in the RoIs.  For our baseline DM profile, we find that the
thermal annihilation cross section can be only reached if instrumental
systematics (namely differential acceptance uncertainties as discussed above)
are under control at the sub-percent level. At the same time, increasing the
time over which the GC is observed by CTA will have a negligible effect on the
projected limits.

\begin{figure}
    \begin{center}
        \includegraphics[width=0.9\linewidth]{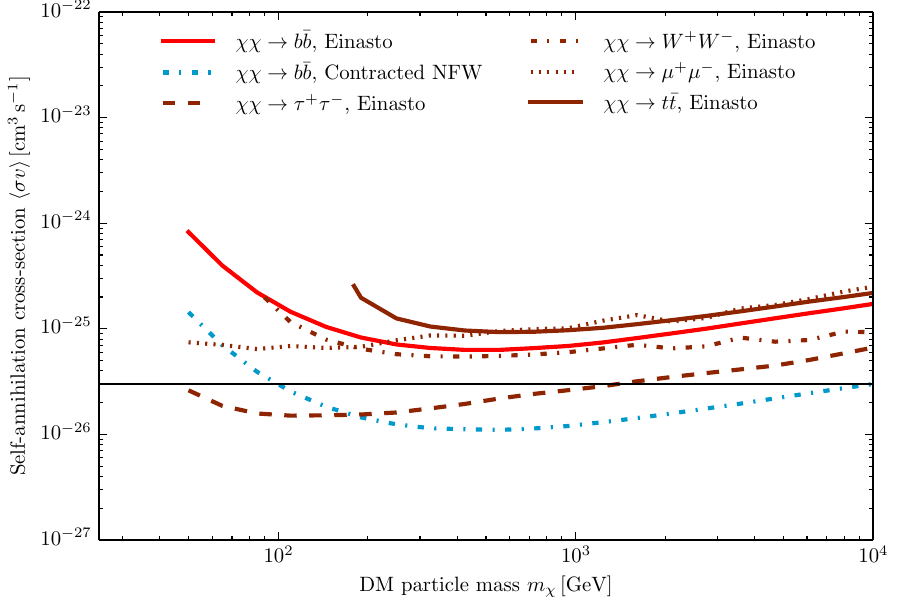}
    \end{center}
    \caption{Comparison of $\langle \sigma v \rangle$ limits from CTA
    observations of the GC, assuming different annihilation channels and DM
    halo profiles. Einasto lines assume the main halo profile described in
    Sec.\ \ref{sec:DarkMatterSignals}.  The contracted NFW profile with an
    inner slope of $\gamma=1.3$ can also be found in that section. All lines
    assume 1\% systematics, 100\,hr of observations and include GDE.}  
    \label{fig:sv_limits_GC_Morpho_ChanProf}
\end{figure}

Finally, we discuss how the projected limits depend on the adopted annihilation
channel or DM halo profile. These results are shown in
Fig.~\ref{fig:sv_limits_GC_Morpho_ChanProf}.  Besides our baseline scenario,
where we assumed an Einasto profile and annihilation into $b\bar{b}$ final
states, we show limits for annihilation to $\tau^+ \tau^-$, $W^+W^-$, $\mu^+
\mu^-$, and $t\bar{t}$ with an Einasto profile, and to $b\bar{b}$ with an alternative density
profile.\footnote{Note that fluctuations of the limits, which are most visible
in the cases of $\mu^+\mu^-$ and $W^+W^-$ final states with strong final state
radiation, come from variations in the adopted effective area.} We find that in
the case of annihilation via the $\tau^+ \tau^-$ channel, CTA would be able to
probe annihilation cross-sections well beyond the thermal value even for a
standard Einasto profile.  In the case of a contracted NFW profile with a inner
slope of $\gamma=1.3$, as described in Sec.~\ref{sec:DarkMatterSignals}, the
$J$-factor increases by a factor of 2.9 (summed over all RoIs).  If this
profile is indeed realised in nature, it would bode well for future
observations with CTA, as CTA could probe well beyond the canonical thermal
cross section for a large range of DM particle masses between 100\,GeV and
10\,TeV.

\section{Conclusion}
\label{sec:conclusion}

In this paper we have performed a new estimate of the CTA sensitivity to DM
annihilation. Here we summarise our main conclusions: 

\begin{itemize} 
    \item {\it We showed that the effect of Galactic diffuse emission
        substantially degrades the sensitivity of CTA when using a
        traditional two region analysis as previous official studies have
        done.} We have assessed the impacts of all backgrounds, including
        protons and electrons in cosmic-rays hitting the atmosphere and, for
        the first time in this type of analysis, diffuse astrophysical
        emission. The impact of including the galactic diffuse emission can be
        observed in particular in Fig.~\ref{fig:diff_sens_GC_OnOff_Fermi},
        where the CTA differential sensitivity is found to be substantially
        degraded (solid line) with respect to the case where the GDE is
        neglected in the analysis (dashed green line). Although we only
        adopted one particular GDE scenario (which is an extrapolation of Fermi
        LAT observations to higher energies), we do not expect these
        conclusions to change when adopting other realistic GDE models.

    \item {\it Including systematic errors substantially degrades the CTA
        sensitivity.} In this paper we introduced a statistical framework that
        allowed us to account for the impact of differential acceptance
        uncertainties within a FoV on DM limits from CTA. This impact can be
        seen in Fig.~\ref{fig:diff_sens_GC_Morpho}, where the sensitivity of
        CTA to DM annihilation is shown for three different values of the
        magnitude of these systematics: 3\% (dashed), 1\% (solid) and 0.3\%
        (dash-dotted).

    \item {\it A morphological analysis substantially improves the sensitivity
        of CTA.} Our morphological analysis allows a proper exploitation of the
        shape differences between the GDE, which is concentrated along the
        Galactic plane, and the DM annihilation signal, which is spherically
        distributed around the GC. The constraints derived under this approach
        are more stringent by a factor of a few compared to those obtained with
        the Ring analysis.  This is best seen by comparing the red
        (morphological) and green (non-morphological) curves in
        Figs.~\ref{fig:diff_sens_GC_Morpho} and
        \ref{fig:sv_limits_GC_OnOff_Morpho}.

    \item {\it Prospects for detecting WIMPs with CTA.} Our most realistic
        estimate of the upper limits on the DM annihilation cross section
        possible with 100\,hr of GC observation by CTA are shown as a red solid
        line in Fig.~\ref{fig:sv_limits_GC_OnOff_Morpho}.  These correspond to
        a morphological analysis assuming annihilation to $b\bar{b}$,
        systematics of 1\%, and include diffuse emission. In order to reach
        canonical thermal cross section, shown as a horizontal line, systematic
        errors should be reduced to less than 0.3\%. If the DM profile is steeper than
        NFW or Einasto, the sensitivity curve drops below the thermal
        cross-section for a broad range of masses, as we show in
        Fig.~\ref{fig:sv_limits_GC_Morpho_ChanProf}.  Here, for the same
        annihilation channel and a contracted profile rising as $r^{-1.3}$, CTA
        is found to be able to probe WIMPs with a thermal cross-section between
        100 GeV and 10 TeV.
\end{itemize}

\paragraph{Acknowledgements.}  We thank Arnim Balzer, David Berge, Jan Conrad,
Christian Farnier, Mathias Pierre and Jennifer Siegal-Gaskins for useful
discussions. PS acknowledges support from the UK Science \& Technology
Facilities Council, and GB from European Research Council through the ERC
Starting Grant {\it WIMPs Kairos}.

\bibliography{CTA_Bibliography,DMbiblio}
\bibliographystyle{JHEP_pat}
\end{document}